\documentclass[twocolumn,showpacs,prb,10pt]{revtex4}
\usepackage[english]{babel}
\usepackage{amsmath,amssymb} 
\usepackage{times}
\usepackage{epsfig} 

\begin{document} 

\title{Phenomenological theory of the anomalous normal-state transport
properties of iron pnictides} 

\author{P. Prelov\v sek$^{1,2}$ and I. Sega$^1$}
\affiliation{$^1$J.\ Stefan Institute, SI-1000 Ljubljana, Slovenia}
\affiliation{$^2$ Faculty of Mathematics and Physics, University of
Ljubljana, SI-1000 Ljubljana, Slovenia}

\date{\today}
\begin{abstract}

We employ the phenomenological theory of the quasiparticle relaxation
based on the simplified two-band description and the spin-fluctuation
mediated interband coupling, to analyze recent normal-state transport data in
electron-doped iron pnictides, in particular the
Ba(Fe$_{1-x}$Co$_x$)$_2$As$_2$ family. The temperature and doping 
dependence of the resistivity, thermopower and the Hall constant are
evaluated. We show that their anomalous behavior emerging from experiments
can be consistently described within the same framework assuming also
``marginal'', i.e., non-Fermi-liquid-like spin fluctuations provided,
however, that the interband  coupling is quite strong. We also show that a large
thermopower as experimentally observed results from an asymmetric energy
dependent quasi particle relaxation rate and due to the semimetallic
character of both bands. 

\end{abstract}

\maketitle 

\section{Introduction}

The novel class of iron-based superconductors (SCs) (Refs. 1 and 2)
reveals besides the high SC transition temperature $T_c$ also several
normal-state properties which are inconsistent with the usual
Fermi-liquid (FL) description of metals. Regarding transport
properties, magnetic ordering and spin fluctuations as well as the
presumable unconventional SC iron pnictides (IPs) are quite close to SC
cuprates.  In the latter class at least part of the anomalous behavior
emerges from the proximity to the Mott-Hubbard insulator \cite{imad}
and from strong correlations, i.e. strong electron-electron Coulomb
repulsion.  On the other hand, in IP correlations seem to be less
severe so the common point might be pronounced low-frequency spin
fluctuations and a strong coupling of charge carriers to such
collective modes.

The experimental evidence for the magnetic order and fluctuations
comes most directly from elastic and inelastic neutron scattering
(INS) showing the commensurate spin-density-wave (SDW), i.e. the
antiferromagnetic (AFM) long-range order in the parent
compounds.\cite{cruz} Recent INS results also confirm strong and
anomalous AFM normal-state spin fluctuations \cite{inos} as well as
the resonant magnetic mode \cite{lums,chi} analogous to the well known
phenomenon in SC cuprates.  That spin fluctuations do not obey normal
FL behavior follows also from NMR-relaxation results.\cite{muku,ning}

The IPs show generally large normal-state d.c. electrical resistivity
$\rho(T)$. Its systematics has been first studied in the family
emerging from the parent (undoped) compound LnFeAsO (LFAO) with a
variety of lantanides Ln=Ce - Dy where the electron doping has been
achieved either by doping with F, e.g., in LnFeAsO$_{1-x}$F$_x$ (LFAO)
\cite{sefa,lee, hess} or via the oxygen deficiency
LnFeAsO$_{1-y}$.\cite{eisa} Generally $\rho(T)$ is very high
comparable to underdoped cuprates.\cite{ando} The behavior changes
from a SDW semimetal $x,y<0.05$ over to the intermediate regime $x,y
\sim 0.1$ with a nearly linear law $\rho \propto T$, into the
overdoped regime with more FL-like $\rho \propto T^2$ behavior for
$y>0.2$. It is not yet evident to what extent very large $\rho(T)$ are
due to polycrystalline character or measured samples since only
recently single-crystal data become available.\cite{jesc} On the other
hand, similar behavior appears in recently studied single-crystal
class of electron-doped (so called 122) family
$AE$(Fe$_{1-x}TM_x)_2$As$_2$ where various alkali elements $AE$=Ba,
Sr, and Ca and transition metals ($TM$=Co,Cu) are at present explored. In
the following we will mostly concentrate on the
Ba(Fe$_{1-x}$Co$_x$)$_x$As$_2$ (BaFeCoAs) compound most investigated
so far where $x$ represents an effective electron doping.  In this
system the qualitative behavior of $\rho(T)$ is similar to LFAO
results although the values are substantially smaller.\cite{fang,rull}
At the same time, the thermopower $S(T)$ is far from FL behavior $S
\propto T$, both in the LFAO (Refs. 10 and 11) and in
BaFeCoAs.\cite{mun} The values become comparable to nondegenerate
electrons, i.e., $|S| \sim s_0=k_B/e_0=86~\mu$V/K with the maximum
typically at $T\sim 100$ K, again resembling underdoped
cuprates.\cite{coop} A similar message is emerging from strongly
$T$-dependent Hall constant $R_H(T)$ in LFAO (Ref. 20) and in
BaFeCoAs.\cite{fang,rull}

Our aim is to extend the phenomenological analysis based on the
simplified two-band model coupled via spin fluctuations as introduced
previously \cite{pst} and apply it for a semi-quantitative description
of the transport properties of the electron-doped IPs, here focusing on
the BaFeCoAs compound. Starting with the evaluation of the anomalous
$T$-dependence of the quasiparticle (QP) damping rates, in particular
in the electron band $\Gamma^e(\omega)$, we analyze within the same
framework the d.c. transport quantities $\rho(T)$, $S(T)$, and
$R_H(T)$. To explain anomalous $\rho(T)$ it is essential to assume
quite strong coupling to non-FL-type spin fluctuations. On the other
hand large $S(T)$ can emerge only via a very asymmetric
$\Gamma^e(\omega)$ and due to the semimetal character (low effective
Fermi energies) of both bands. The previous approach \cite{pst} is
here upgraded with an explicit evaluation of $T$-dependent
$\Gamma^e(\omega)$ and corresponding d.c. transport quantities without
further simplifications. Qualitatively one can follow the development
from a non-FL behavior to a more normal FL-type regime by changing the
carrier concentration and the character of the input spin fluctuations by
introducing a crossover temperature $T^*$ controlling the extent of non-FL
vs. normal FL behavior thus simulating the experimental findings in
BaFeCoAs as well as in the LFAO system.

In Sec.~II we describe the phenomenological model as introduced
previously \cite{pst} and give some justification based on a more
complete microscopic model. Section~III presents the lowest order
approximation for the QP damping $\Gamma(\omega)$ being the central
ingredient for the understanding of the transport properties where its 
temperature and energy dependence are essentially controlled by the
(phenomenological) ansatz for the spin-fluctuations mediated interband
coupling as noted above. In Sec.~IV basic equations for the d.c. transport
quantities  are presented. Results in different regimes of relevant
parameters follow in Sec.~V with a discussion of the approach and results in
Sec.~VI. 

\section{Two-band model}

For the present analysis we adopt a simplified 2D model for IP (Ref. 21) 
taking into account only two bands, an electron (e) band and a hole (h)
band, both crossing the Fermi surface \cite{mazi,ragh,chen} and coupled via
spin fluctuations.\cite{pst} In the folded Brillouin zone
\cite{kors,mazi} the h-like and e-like pockets are at ${\bf k} \sim
0$, and ${\bf k} \sim {\bf Q} =(\pi,\pi)$, respectively,
\begin{eqnarray}
H_{ef}&=&\!-\!\sum_{{\bf k},s}
\bigl(\zeta^e_{\bf k} c^\dagger_{{\bf k}s} c_{{\bf k}s} + \zeta^h_{\bf
  k} d^\dagger_{{\bf k}s} d_{{\bf k}s} \bigr) \nonumber \\ &+&
\frac{1}{\sqrt{N}} \sum_{{\bf kq},ss^\prime} m_{\bf kq} {\bf S}_{\bf
  q}\cdot {\bf \sigma}_{ss\prime}c^\dagger_{{\bf k}-{\bf q},s}
d_{{\bf k}s'}+ H.c.), \label{mod}
\end{eqnarray}
and $c_{\bf k},d_{\bf k}$, ($\zeta^e$, $\zeta^h$) refer to
electrons in e-like and h-like bands, respectively.

The following justification can be given for the above
phenomenological model. The realistic electronic model for 2D IPs
includes several (e.g., five) bands, emerging from $d$ orbitals of
Fe. The interaction term is very complicated in general. Still the
low-energy description (in the case of weak or modest
electron-electron interactions) should contain only both bands at the
Fermi surface (FS). So one of the relevant interaction terms
generating the interband coupling in Eq.(\ref{mod}) can be written as
\cite{chub}
\begin{equation}
\tilde H = \sum\tilde U c^\dagger_{{\bf k}s} d^\dagger_{{\bf k'}s'}
d_{{\bf k''}s'} c_{{\bf k'''}s}. \label{int}
\end{equation}
The origin of the above interaction resides within a more complete
multi-orbital model, the inter-orbital and intra-orbital Hubbard
interactions $U$ and $V$, respectively, as well as in the Hund's coupling
$J_H$. The effective interaction, Eq.(\ref{int}), could be rewritten
as the coupling to a fluctuating SDW field
\begin{equation}
{\bf {\cal S}}_{\bf q} = \tilde U \sum_{{\bf k}ss'} {\bf \sigma}_{ss'}
c^\dagger_{{\bf k}s} d_{{\bf k+q}s'}.
\end{equation}
The SDW fluctuation ${\bf{\cal S}}_{\bf q} $ should have in general,
e.g., for the relevant ${\bf q} \sim {\bf Q}$ a finite projection on the
usual spin operator ${\bf S}_{\bf q}$ (emerging from local moments),
but this remains to be shown by an explicit consideration within the
full multi-orbital model for IPs.

\section{Quasiparticle damping}

In the evaluation of the transport properties we follow the approach
introduced previously.\cite{pst} We first consider the corresponding
Green's functions for $\sigma =e,h$ $\langle\bar \sigma=(h,e)\rangle$ electrons
$G_{\bf k}^\sigma(\omega)= [\omega^+ - \epsilon^\sigma_{\bf k} -
  \Sigma^{\sigma}_{\bf k}(\omega)]^{-1}$, where $\epsilon^\sigma_{\bf
  k} =\zeta^\sigma_{\bf k} - \mu$. The self-energies
$\Sigma^{\sigma}_{\bf k}(\omega)$ are evaluated within the
lowest-order perturbation in the interband coupling, Eq.(\ref{mod}),
induced by spin fluctuations entering via dynamical spin
susceptibility $\chi_{\bf q}(\omega)$,
\begin{eqnarray}
\Sigma^{\sigma}_{\bf k}(\omega)\!\!\!\!&&=3 \sum_{\bf q} m^2_{\bf k q} \int\!
\!\int \frac{d\omega_1 d\omega_2}{\pi} g_{12} \frac{A^{\bar\sigma}_{{\bf k}-{\bf
q}} (\omega_1) \chi''_{\bf q}(\omega_2)}
{\omega-\omega_1-\omega_2 }, \nonumber \\
g_{12}\!\!\!\!&&\equiv g(\omega_1,\omega_2)= \frac{1}{2}\bigl [{\rm
th}\frac{\beta\omega_1}{2}+{\rm cth}\frac{\beta\omega_2}{2} \bigr]. 
\label{sig}  
\end{eqnarray}
To proceed we make several simplifications, which are expected to
apply to IPs in the low-doping regime close to the AFM (in the folded
zone) instability, i.e., in the system without long-range AFM
order. The spin fluctuations $\chi_{\bf q}''(\omega)$ in the normal
phase are assumed to be centered at ${\bf q} \sim {\bf Q} = (\pi,\pi)$
and broad enough in the ${\bf q}$ space relative to h/e pockets. This
is indeed well visible in recent INS results for the BaFeCoAs
system.\cite{inos} Hence we replace $\chi_{\bf q}(\omega) \sim
\chi_{\bf Q}(\omega) =\tilde \chi(\omega)$. Since the relevant e/h
bands form only small pocket-like FS we neglect in the low $\omega$
regime also the ${\bf k}$ dependences of the self-energies; i.e., we
replace $\Sigma^\sigma_{\bf k}(\omega) \sim \Sigma^\sigma(\omega)$
whereby ${\bf k} \sim 0$ and ${\bf k} \sim {\bf Q}$ are relevant for
the h- and e-bands, respectively. Hence, with the local character of
the self energies the spectral function $A_{\bf  k}^\sigma(\omega) =
A^\sigma(\epsilon_{\bf k},\omega)$ can be expressed as
\begin{equation}
 A^\sigma(\epsilon_{\bf k},\omega)=- \frac{1}{\pi} \mathrm
 {Im}[\Omega^\sigma(\omega) - \epsilon^\sigma_{\bf k}
   +i\Gamma^\sigma(\omega)]^{-1}, \label{akom}
\end{equation}
with
$\Omega^\sigma(\omega)=\omega-\mathrm{Re}\Sigma^\sigma(\omega)=\omega/
Z^\sigma(\omega)$ and
$\Gamma^\sigma(\omega)=-\mathrm{Im}\Sigma^{\sigma}(\omega)$. Note that
$Z^\sigma(\omega)$ and $\Gamma^\sigma(\omega)$ play the role of the QP
weight and the QP damping, respectively.
  
The above approximations simplify the expression for the QP damping as it
follows from Eq.(\ref{sig}),\cite{pst}
\begin{equation}
\Gamma^{\sigma}(\omega)=
\frac{3}{2}  \bar m^2 \int d\omega' g(\omega-\omega',\omega') {\cal
  N}^{\bar\sigma}(\omega-\omega') \tilde \chi''(\omega'), \label{gam}
\end{equation}
where the effective interband coupling is $\bar m \sim m_{\bf Q,Q} =
m_{\bf 0,Q} $ and ${\cal N}^{\sigma}(\omega)=(2/N)\sum A^\sigma_{\bf
  k}(\omega)$ are the e/h band single-particle densities of states (DOS).

Within the present analysis spin fluctuations are taken as a
phenomenological input. In analogy with the anomalous QP and transport
properties of cuprates \cite{imad} we assume that the origin is in the
non-FL character of $\chi''_{\bf q}(\omega)$ and a strong coupling of
carriers to the latter. Indeed, recent INS results in electron-doped
BaFeCoAs indicate a behavior very close to the one for marginal
FL.\cite{inos} For generality we therefore assume further-on the form
\begin{equation}
\tilde \chi''(\omega)=\pi C(\omega) {\rm th}\frac{\omega}
{2(T+T^*)}, \label{chi}
\end{equation}
where $C(\omega)$ is (for $T^*=0$) the (symmetrized) dynamical spin
correlation function. While $T^*=0$ yields the marginal FL (Ref. 27) 
dynamical fluctuations, we can simulate with $T^*>0$ the transition to
the normal FL regime for low $T<T^*$. For simplicity also a smooth
cutoff on $C(\omega)=C_0$ is imposed for $\omega>\omega^*$.

\section{DC transport quantities}

dc electrical conductivity $\sigma_0$ and the thermopower $S$ can be
expressed via general transport coefficients. For the isotropic
transport in 2D and assuming separate contributions of both bands,
transport coefficients $L_{1n}^\sigma$ can be expressed within the
linear response theory in terms of spectral functions \cite{maha}
provided also that vertex corrections are neglected,
\begin{equation}
L^\sigma_{1n}= -\frac{2 \pi}{N} \sum_{\bf k}
(v_{\bf k}^{\alpha\sigma})^2 \int d\omega \omega^{n-1} \frac{df}{d\omega}
 A^\sigma_{\bf k}(\omega), \label{ln}
\end{equation}
with the Fermi function $f=1/[{\rm exp}(\omega/T)+1]$ and 
$v_{\bf k}^{\alpha\sigma}$ $\sigma$-band velocities ($\alpha=x,y$). Due to the
${\bf k}$-independent $\Sigma^\sigma(\omega)$ and Eq.(\ref{akom}) the actual
expressions have the same form as used within the dynamical-mean-field
theory,\cite{pals}
\begin{eqnarray}
L^\sigma_{1n} =-\pi \int \int d\omega d\epsilon \phi^\sigma(\epsilon)
\omega^{n-1} \frac{df}{d\omega}[ A^\sigma(\epsilon,\omega)]^2, \nonumber \\
\phi^\sigma(\epsilon)=\frac{2}{N} \sum_{\bf k} (v^{\alpha\sigma}_{\bf k})^2
\delta(\epsilon-\epsilon_{\bf k}), \label{ln1}
\end{eqnarray}
With known $L^\sigma_{1n}$ one can express the final d.c. conductivity
\begin{equation}
\sigma_0=e_0^2L_{11}=e_0^2(L^e_{11}+L^h_{11}), \label{sig0}
\end{equation}
as well as the thermopower
\begin{equation}
S_0=-s_0 \frac{L_{12}}{T L_{11}}, \label{seebeck}
\end{equation}
where $L_{12}=L_{12}^e+L_{12}^h$ and $s_0=k_B/e_0$.

Within the same framework and approximations an analogous
expression for the Hall conductivity has also been derived
\cite{kohno,voru} and applied to nontrivial models \cite{meri,haul}
\begin{equation}
\sigma^\sigma_{xy}= -\sigma_0^H \int \int d\omega d\epsilon
\phi_H^\sigma(\epsilon)  \frac{df}{d\omega}[A^\sigma(\epsilon,\omega)]^3, 
\label{sxy}
\end{equation}
where $\sigma_0^H=2\pi^2 e_0^3 B/3$ and
\begin{equation}
\phi_H^\sigma=\frac{2}{N} \sum_{\bf k}\big(\frac{\partial
  \epsilon^\sigma_{\bf k}}{\partial k_x}\big)^2 \frac{\partial^2
  \epsilon^\sigma_{\bf k}} {\partial k_y^2}
\delta(\epsilon-\epsilon^\sigma_{\bf k}). \label{phih}
\end{equation}
Again, within the two-band model
$\sigma_{xy}=\sigma^e_{xy}+\sigma^h_{xy}$ and
\begin{equation}
R_H= \frac{\sigma_{xy}}{B \sigma_0^2}. \label{rh}
\end{equation}
It should be, however, recognized that Eq.(\ref{sxy}) obtained via the
linearization of the linear response in an external magnetic field
$B>0$ is less explored (as compared with transport coefficients
Eq.(\ref{ln})) and only partly tested in situations with nontrivial
$\Gamma^\sigma(\omega) \neq \Gamma^\sigma_0$, in particular in the
presence of correlations.\cite{kont}

Transport quantities depend only on the QP properties close to the Fermi
surface. Hence, we assume in the following calculations the simplified
form for the unperturbed e/h bands, i.e. with 2D parabolic dispersions,
\begin{equation}
\epsilon^e_{\bf k}=\epsilon^e_0+ t_e k^2,
\quad \epsilon^h_{\bf k}=\epsilon^h_0- t_h k^2, \label{ek}
\end{equation}
with (unrenormalized) effective masses $m^\sigma=1/(2t^\sigma)$ (with
lattice constant $a_0=1$) and the unperturbed DOS 
${\cal N}^\sigma_0=1/(2\pi t^\sigma)$. Corresponding $\phi^\sigma,
\phi_H^\sigma$ functions follow from Eqs.(\ref{ln1}) and (\ref{phih}),
\begin{equation}
\phi^\sigma(\epsilon) =2  t^\sigma |\epsilon-\epsilon^\sigma_0|
{\cal N}^\sigma_0, \quad
\phi_H^\sigma(\epsilon) = \pm 2 t^\sigma \phi^\sigma(\epsilon).
\end{equation}
On the other hand, the final DOS entering Eq.(\ref{gam}) involves the
effects of nontrivial $\Sigma^\sigma(\omega)$ and is evaluated as
\begin{eqnarray} 
&&{\cal N}^\sigma(\omega) = -\frac{2}{\pi W^\sigma}{\rm Im} \int
\frac{d\epsilon}{\Omega(\omega) -\epsilon +i \Gamma(\omega) } = 
\nonumber \\ &&\frac{2}{\pi W^\sigma} \bigl[ {\rm
    arctg}\frac{\Omega^\sigma-\epsilon^\sigma_0+W^\sigma}{\Gamma^\sigma}
  -{\rm arctg}\frac{\Omega^\sigma-\epsilon^\sigma_0}{\Gamma^\sigma}
  \bigr],
\label{dos}
\end{eqnarray}
where $W^\sigma=4\pi t^\sigma$ is the bandwidth within the parabolic
approximation. In addition, we assume that the densities of electrons
within the e/h bands, i.e. $n_e=x_e$ and $n_h=1-x_h$, $x_h$ being
the density of holes, are fixed separately, so that at each $T>0$ the
sum rule
\begin{equation}
n^\sigma= \int d\omega f(\omega) {\cal N}^\sigma(\omega),\label{nsig}
\end{equation}
should be satisfied leading in principle to a nontrivial shift of
$\epsilon^\sigma_0(T)$ in Eq.(\ref{ek}) (with respect to the chemical
potential).

\section{Results}

The above equations fully determine the evaluation procedure for
$\Sigma^\sigma(\omega)$ at any $T$ and consequently of transport quantities
$\rho_0(T), S(T), R_H(T)$. The main input parameters are the e/h
concentrations $x_e, x_h$ and the corresponding $t^\sigma$. We note that
coupling to spin fluctuations has a meaningful dimensionless
coupling constant
\begin{equation}
g_0= \bar m^2 C_0 \sqrt{{\cal N}_0^h{\cal N}_0^e}. \label{g0}
\end{equation}
In addition, results depend also on the FL temperature $T^*$ in
Eq.(\ref{chi}) and on the cutoff $\omega^*$, the latter mainly
influencing low-$\omega$ behavior through the QP weight $Z(\omega)$.

In the following we consider the electron doped case $x_e>x_h$, as
relevant, e.g. for the electron-doped BaFeCoAs. Obviously, here the
transport properties are dominated by the transport within the e-type
band, although at high $T$ also hole pockets contribute partially.  On
the other hand, we realize that in such a case the treatment of
minority (h) carriers is oversimplified, since even at modest $g_0 <
1$ the hole QPs become heavily renormalized $Z^h(\omega) \ll 1$
influencing also ${\cal N}^h(\omega)$. The latter could be an artifact
of the lowest order approximation for the QP damping,
Eq.(\ref{gam}). To avoid rather unphysical properties of hole QP, we
introduce for the latter a simplification
$\Gamma^h(\omega)=\Gamma^h_o$.  Note that $\Gamma^h$ influences
results only indirectly, i.e. entering the dominant e-band damping
$\Gamma^e(\omega)$ via ${\cal N}^h(\omega)$. At the same time, the
hole contribution to d.c. transport quantities remains subleading
provided that $\Gamma^h_0>\Gamma^e(\omega \sim 0)$, in particular at
low $T$.

Our aim is to show that the presented phenomenological theory can
consistently as well as semi-quantitatively account for the anomalous
non-FL normal-state transport properties of (electron-doped) IPs, in
particular for the large and $T$-linear $\rho(T)$, large thermopower
$S(T)$ reaching quasiclassical values $|S| \sim s_0$, and $T$-dependent
$R_H(T)$. Evidently, in the absence of a better understood
microscopic model and relevant input parameters as the effective spin
fluctuations, all varying between materials and different doping
levels, we here mainly consider various possibilities and regimes
which emerge from our analysis.

While displaying different regimes we fix some parameters. In the
following we choose (unrenormalized) band parameters $t_e=0.4$~eV and
$t_h=0.3$~eV and spin-fluctuation cutoff $\omega^*=0.5$~eV. Actual
levels of carrier concentrations $x_e,x_h$ are not directly known in
particular materials, since the doping $x$, e.g., with Co in BaFe$TM$As,
generally gives only partial information on $x=x_e-x_h$. To explain
the anomalous properties it is, however, essential that $x_h$ is low
leading to a very asymmetric damping $\Gamma^e(\omega)$ and
consequently to large values of $S(T)$. On the other hand, a
disappearance of $x_h \to 0$ reduces strongly the scattering within the
e-band and the transport is expected to become more FL-like.

Evidently, the crucial parameter is the coupling strength $g_0$.  It
most directly influences the resistivity $\rho(T)$, and in particular
its derivative $d\rho(T)/dT$. To account quantitatively for $T\sim
300$~K experimental results for BaFeCoAs, we require modest $g_0\sim{\cal O}(1)$
as discussed further on.  On the other hand, much larger resistivities
within the LFAO family indicate substantially larger $g_0 \gg 1$, as
analyzed previously.\cite{pst} It should be noted that due to
constant $\Gamma^h_0$ and $x_h$ iterations of Eq.~(\ref{sig}) are not
needed but still $\mu$ should for each $T$ separately be determined to
keep $x_e$ fixed.
  
\begin{figure}[htb]
\includegraphics[angle=0, width=.8\linewidth]{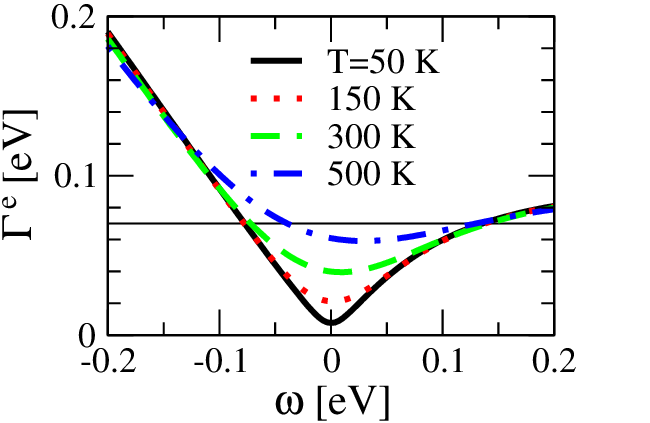}
\caption{(Color online) Electron-band QP damping 
$\Gamma^e(\omega)$ for various $T$ for the intermediate doping.
Horizontal line represents the chosen hole damping $\Gamma^h_0$.}
\label{fig1}
\end{figure}

\begin{figure}[htb]
\includegraphics[angle=0, width=.8\linewidth]{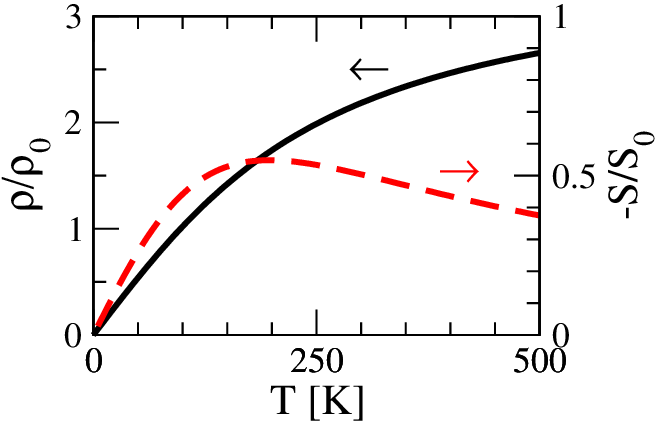}
\caption{(Color online) The resistivity $\rho(T)/\rho_0$ (ful line, left
scale) and the thermopower $S(T)/S_0$ (dashed line, right scale) for the
intermediate electron-doped IP.} 
\label{fig2}
\end{figure}

Let us start with the case, representing presumably the anomalous
intermediate doping. For actual results presented in Figs.~1 and 2 we
choose fixed concentrations $x_e=0.2, x_h=0.08$ while the remaining
parameters are: $g_0=1.6$, $T^*=0$, and $\Gamma^h_0=0.07$~eV.  In Fig.~1 we
first present the variation of $\Gamma^e(\omega)$ around $\omega \sim
0$ for various $T$. Since $T^*=0$ the general behavior is of the
MFL-type, i.e. roughly $\Gamma^e \propto {\rm  max}(|\omega|,T)$.\cite{pst}
But as well important is a pronounced 
asymmetry with respect to $\omega=0$ which emerges from low $x_h \ll
1$ and from the strong scattering on $\omega$-dependent ${\cal  N}^h$. The
corresponding $\rho(T)/\rho_0$ and $S(T)/S_0$ are 
presented in Fig.~2. Note that within a 2D layer
$\rho_0=\hbar/e_0^2=41.5$~k$\Omega$. For actual BaFeCoAs materials with the
interlayer distance $c_0\sim   1.3$~nm this corresponds to
$\tilde\rho_0=c_0\rho_0\sim~0.5$~m$\Omega$cm. As a consequence of the MFL-type
$\Gamma^e(\omega)$ also $\rho(T)$ displays a linear variation for
lower $T<300$~K with a tendency to a saturation at higher $T$. Even more
anomalous is $S(T)$ reaching a minimum at low $T \sim 150$~K with
anomalously large (for a metal) value $S \sim -0.5 s_0$. The decrease
in $|S|$ at higher $T$ results mainly from the compensation effect
between e-band and h-band contributions.

The effect of a nonzero FL temperature $T^*$ in Eq.(\ref{chi}), is
presented in Fig.~3. At $T^*=0$ the MFL-type variation is evident in
the linear resistivity $\rho \propto T$ in Fig~3a. Finite $T^*>0$
induces more normal FL-type variation $\rho \propto T^2$ for
$T<T^*$. At the same time also the $S(T)$ changes gradually from an
anomalously large and non-monotonous variation at $T^*=0$ into more
normal FL-type $S\propto T$ for large $T^*=1000$~K. The Hall constant
$r_H(T)=R_H(T)/R_{H}^0$ (where $R_{H}^0=V_0/e^2$, $V_0$ being the
volume of the unit cell) is less sensitive on the non-FL
behavior. Still it shows pronounced $T$ variation for $T^*=0$ emerging
from a nontrivial electron/hole compensation and linear $\rho_e(T)
\propto T$. Finally, at large $T^*$  one sees in Fig.~3c rather constant
$r_H(T) \propto -1/x_e$ as appropriate for a normal metal.

\begin{figure}[htb]
\includegraphics[angle=0, width=.7\linewidth]{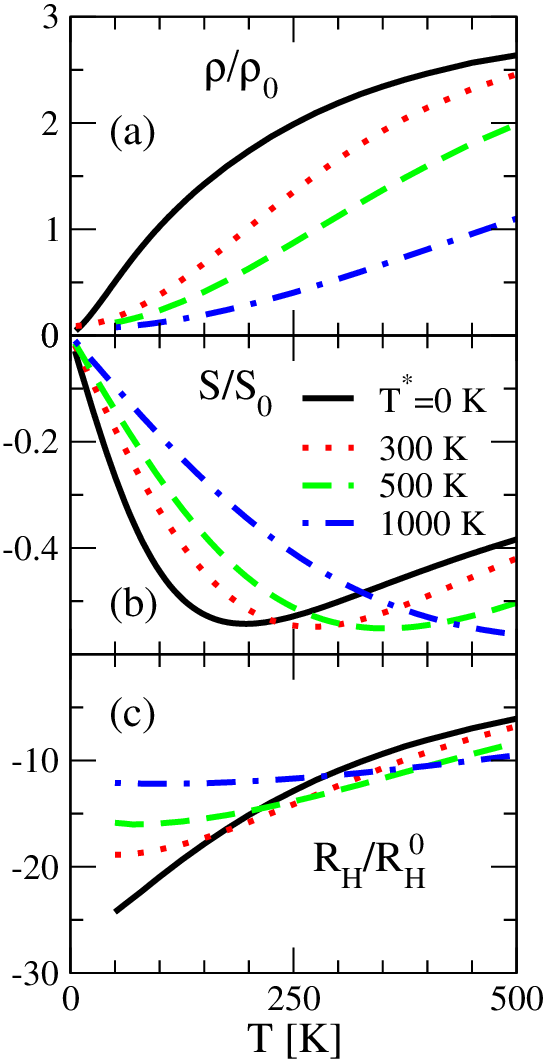}
\caption{(Color online) a) Resistivity $\rho(T)/\rho_0$, b)
  thermopower $S(T)/S_0$, and c) the Hall constant $R_H(T)/R_H^0$ for
  different Fermi-liquid temperatures $T^*$ and fixed concentrations
  $x_e=0.2, x_h=0.08$. }
\label{fig3}
\end{figure}

As last, we discuss the variation with the electron doping $\delta
n_e$. We assume as the parent system the semimetal with finite
$x_e^0=x_h^0=0.15$. The effect of doping leads then to an increase in
$x_e$ where we take for simplicity $x_e=x^0_e+\delta n_e/2$ and
accordingly $x_h=x_e-\delta n_e$.  Since our treatment with a
constant $\Gamma^h_0$ and without the possibility of an AFM ordering
is not adapted for an undoped system, we consider here only a limited
range $\delta n_e = 0.1 - 0.2$. The change in $\rho(T)$ in Fig.~4a is
quite dramatic at $\delta n_e=0.2$. This is a direct consequence of
nearly vanishing $x_h \sim 0$ where the interband scattering becomes
ineffective at low $\omega$. Such a situation in real IPs corresponds
to the well overdoped case. Clearly, our simplified analysis becomes
insufficient beyond this point. Also $S(T)$ shows a similar trend, since
at highest doping it becomes entirely FL-like with $S = \alpha T$
with a modest $\alpha$.

\begin{figure}[htb]
\includegraphics[angle=0, width=.7\linewidth]{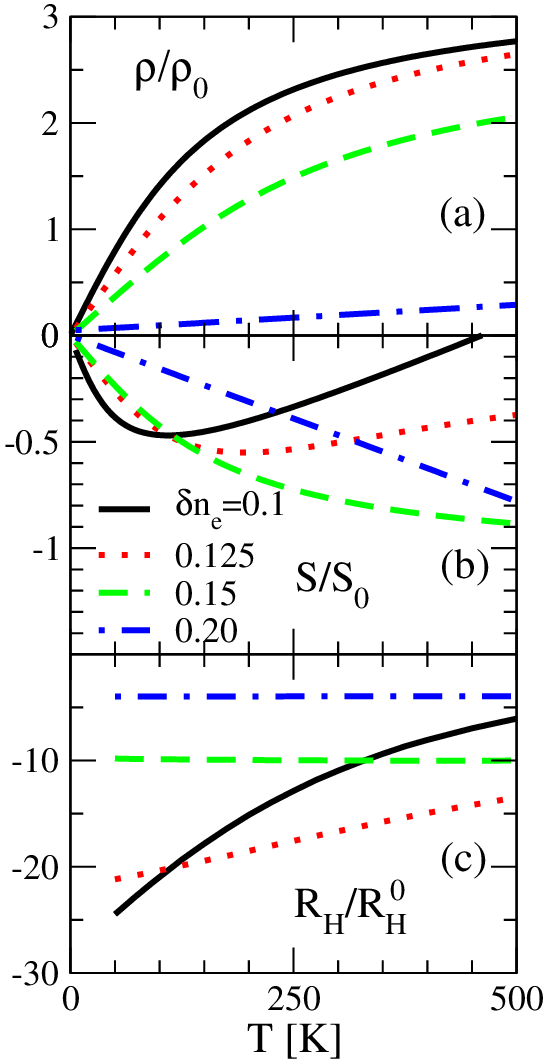}
\caption{(Color online) a) Resistivity $\rho(T)/\rho_0$, b)
thermopower $S(T)/S_0$, and c) the Hall constant $R_H(T)/R_H^0$ for
$T^*=0$ and various electron doping $\delta n_e$ with respect to the
parent $x_e^0=x_h^0=0.15$ case. }
\label{fig4}
\end{figure}

Finally, let us turn to the comparison of our results with the
experimental data on electron-doped IPs. Due to the phenomenological
character of our theory as well as different IP material classes with
quite different properties only a semi-quantitative comparison makes
sense. Our central goal is to reproduce quite systematic variation of
transport properties with doping, from low electron doping over the
intermediate regime (optimal doping with highest $T_c$) to the
overdoped one where SC disappears and IPs become rather normal FL-type
metals. In our approach the doping dependence enters primarily through
the electron/hole concentration difference $\delta n_e$.  But as well,
it is plausible that also the spin-fluctuation spectra
$\tilde\chi''(\omega)$ change, being strong and anomalous at low
doping, and on the other hand at intermediate doping weaker and more
FL-type in the low-$\omega$ regime. Within our analysis we can partly
simulate the latter development with increasing $T^*$.

In the most challenging intermediate-doping regime our transport
results match reasonably recent experiments on single-crystals of the
Co-doped BaFe$_2$As$_2$ compound. Note that BaFeCoAs with $x=0.1$ at $T=300$~K
typically reveal $\rho \sim 0.3$~m$\Omega$cm with a rather $T$-linear
variation and corresponding slope $d\rho/dT \sim 0.5$~$\mu\Omega$cm/K.
\cite{fang,mun,rull} As shown in Fig.~2a the
value (note that $\tilde\rho_0 = 0.5$~m$\Omega$cm) and the slope are
reproduced within a factor of $3-4$. While the magnitude of $\rho$ is
mostly governed by the spin-fermion coupling $g_0$ and can be fitted
to experiments accordingly, it is essential that our analysis also
reproduces the anomalous variation of $S(T)$. Recent data on BaFeCoAs
at $x \sim 0.15$ reveal an abrupt increase in electron-like
thermopower with a maximum value $|S| \sim 0.6 s_0$ at $T =T_0 \sim
100$~K.\cite{mun} For $T>T_0$ $|S|$ is decreasing in value. Both these
features as well as values are well reproduced in Fig.~2b.

With increasing Co-doping in BaFeCoAs, $x>0.1$, the resistivity
behavior becomes closer to $\rho(T) \propto T^2$ not so much reduced
in value at $T \sim 300$~K.\cite{fang,rull} In our analysis this
corresponds to a change obtained by increasing the FL temperature
$T^*>0$ with a simultaneous increase in $\delta n_e$ as presented in
Figs.~3a and 4a. Even more pronounced change in character is in $S(T)$
(Ref. 18) since it reduces in value, loses the minimum and becomes
FL-like, i.e., $S=\alpha T$. Such a development is well visible in
Figs.~3b and 4b since both increasing $T^*>0$ and $\delta n_e$ lead to a
crossover to a normal FL with a moderate slope $\alpha$.

Experiments on the Hall constant reveal an electron-like $R_H<0$ with a
pronounced $T$ dependence \cite{mun,rull} which reduces in the
overdoped regime where at the same time it becomes consistent with the
quasiclassical picture $R_H/R^0_H \sim -1/x_e$.  Our results in
Figs.~3c and 4c show that both $T^*>0$ as well as $\delta n_e > 0$ lead to
a constant FL-like $R_H$. The particular value at $T \to 0$ from our
analysis is less reliable. While at weak coupling we recover the
expected $R_H/R^0_H \sim -1/x_e$ (for $\rho_e \ll \rho_h$) the value
can deviate for larger coupling since the approximations do not
satisfy the Luttinger-volume sum rule. Note, however, that also the
basic approximation Eq.(\ref{sxy}) is not well understood
yet. \cite{voru,haul}

Qualitatively similar results for transport quantities have been
previously obtained for polycrystalline samples of the electron-doped
LFAO family \cite{hess}, discussed previously within the same
phenomenological framework.\cite{pst} While our present data with
assumed parameters would quite well match the measured
$S(T)$,\cite{sefa} reported $\rho(T)$ as well as the slope $d\rho/dT$
(Ref. 12) are higher than in BaFeCoAs by a factor $\sim 5$. It seems
indicative that also recent single-crystal data \cite{jesc} do not
reduce these values. From our present analysis this would require
larger $g_0 \gg 1$ which could open the question of entering the (too)
strong coupling regime.

\section{Discussion}

In conclusion, let us first put the present theoretical approach and
its possible extensions in the perspective of related theoretical
approaches, in particular those treating the QP, magnetic, and
transport quantities in IPs. In contrast to many attempts to start with
a more complete microscopic model,\cite{mazi} our approach is
restricted to the simplest prototype two-band model coupled via spin
fluctuations. These are taken as a phenomenological input of the
marginal FL type whose deviation from the normal FL behavior is controlled
by the crossover temperature $T^*$. Our analysis clearly indicates that the
coupling appears not to be weak $[$at least as deduced from observed
$\rho(T)$$]$ and that spin fluctuations are quite strong in the low-$\omega$
regime as evidenced by INS experiments.\cite{inos} It is therefore expected
that our analysis can go beyond the weak-coupling approximation, being the 
basis, e.g., for a random-phase-approximation-type treatment of many-orbital models. The
latter has been tested so far mostly on reproducing the proper
band structure and on a qualitative description of relevant
normal-state instabilities, in particular the most challenging SC
onset. Evidently, our approach goes beyond the simple nesting scenario
of the QP damping as well as of the SC pairing since in the latter wave
vector (${\bf k},{\bf q}$) dependences would play an essential role.

Interpreting our results for anomalous transport properties of IPs, we
argue that they emerge as the interplay of several ingredients. First
of all, observed $\rho(T)$ requires large interband coupling since the
transport becomes quite normal and modest when the electron doping
depletes the hole band, i.e. on approaching $x_h \sim 0$.  The magnetic
origin of the coupling is quite natural since it is intimately
connected to the AFM (SDW) instability in the parent material and
leads to an analogy with the physics of SC cuprates.

Highly nontrivial $S(T)$ in our analysis can be mostly related to two
effects: a) asymmetric variation of the QP damping $\Gamma^e(\omega)$ being
the consequence of asymmetric DOS ${\cal N}^h(\omega)$ due to the closeness
of the top of the hole band, and b) small $\epsilon_e^0$ which plays a role
of the electron Fermi energy entering directly the expected FL behavior $S
\sim \pi^2 T/3 \epsilon_e^0$.         

There are still several open questions even within the presented
framework.  Generally, stronger interband coupling leads to strong
damping of minority QP which requires the treatment beyond the weak
coupling. The same obstacle arises in the treatment of the parent and a
weakly doped semimetal, where also experiments indicate very
incoherent transport above $T>T_c$. Also, a comparison with
experimental results for particular compounds could lead to refined
models to reach a closer quantitative agreement.

\section{Acknowledgments}

One of us (P.P.) acknowledges the fruitful discussion with T.
Tohyama. This work was supported in part by the Slovenian Agency for
Research and Development and by MHEST and JPSJ under the Slovenia-Japan
Research Cooperative Program.


\begin{thebibliography}{99}
\bibitem{kami} Y.\ Kamihara, T.\ Watanabe, M.\ Hirano, and H.\ Hosono,
  J. Am. Chem. Soc. {\bf 130}, 3296 (2008).

\bibitem{hoso} for a review see H.\ Hosono and Z.-A.\ Ren, New
  J. Phys. {\bf 11}, 025003 (2009); K.\ Ishida, Y.\ Nakai, and H.\ Hosono,
  J.\ Phys.\ Soc. Jpn. {\bf 78}, 062001 (2009).

\bibitem{imad} M.\ Imada, A.\ Fujimori, and Y.\ Tokura, 
Rev. Mod. Phys. {\bf 70}, 1039 (1998).

\bibitem{cruz} C.\ de\ la\ Cruz {\it et al.}, Nature {\bf 453}, 899 (2008).

\bibitem{inos} D.\ S.\ Inosov {\it et al.}, arXiv:0907.3632v1.

\bibitem{lums} M.\ D.\ Lumsden {\it et al.}, Phys. Rev. Lett. {\bf 102},
107005 (2009). 
  
\bibitem{chi} S.\ Chi {\it et al.}, Phys. Rev. Lett. {\bf 102}, 107006 (2009).

\bibitem{muku} H.\ Mukuda {\it et al.}, J. Phys. Soc. Jpn. {\bf 78}, 084717
(2009).

\bibitem{ning} F.\ L.\ Ning {\it et al.}, Phys. Rev. Lett. {\bf 104}, 037001
(2010).

\bibitem{sefa} A.\ S.\ Sefat {\it et al.}, Phys. Rev. B{\bf 77}, 174503 (2008).

\bibitem{lee} S.\ C.\ Lee {\it et al.}, J. Phys. Soc. Jpn., {\bf 78}, 043703
(2009). 

\bibitem{hess} C.\ Hess {\it et al.}, Europhys. Lett. {\bf 87}, 17005 (2009). 

\bibitem{eisa} H.\ Eisaki {\it et al.}, J. Phys. Soc. Jpn. {\bf 77}, Suppl.
C, 36 (2008); K.\ Miyazawa {\it et al.}, J. Phys. Soc. Jpn. {\bf 78}, 034712
(2009).
\bibitem{ando} Y.\ Ando, A.\ N.\ Lavrov, S.\ Komiya, K.\ Segawa, and X.\ F.\
Sun, Phys. Rev. Lett. {\bf 87}, 017001 (2001); Y.\ Ando, S.\ Komiya, K.\
Segawa, S.\ Ono, and Y.\ Kurita, Phys. Rev. Lett. {\bf 93}, 267001 (2004). 

\bibitem{jesc} A.\ Jesche, C. Krellner, M.\ de Souza, M.\ Lang, 
and C.\ Geibel, Ne. J. Phys. {\bf 11}, 103050 (2009).
  
\bibitem{fang} L.\ Fang {\it et al.}, Phys. rev. B{\bf 80}, 140508(R) (2009).

\bibitem{rull} F.\ Rullier-Albenque, D.\ Colson, A.\ Forget, and H.\ Alloul,
Phys. Rev. Lett. {\bf 103}, 057001 (2009).

\bibitem{mun} E.\ D.\ Mun, S.\ L.\ Budko, N.\ Ni, and P.\ C.\ Canfield,
Phys. Rev. B{\bf 80},054517 (2009..  

\bibitem{coop} J.\ R.\ Cooper and J.\ W.\ Loram, J. Phys. I (France),
{\bf 6}, 2237 (1996).

\bibitem{koha} Y.\ Kohama {\it et al.}, Europhys. Lett. {\bf 84}, 37005
  (2008).

\bibitem{pst} P.\ Prelov\v sek, I.\ Sega, and T.\ Tohyama,
Phys. Rev. B{\bf 80}, 014517 (2009).

\bibitem{mazi} for a theoretical overview see I.\ I.\ Mazin and
  J.\ Schmalian, arXiv:0901.4790v1.

\bibitem{ragh} S.\ Raghu, X.-L.\ Qi, C.-X.\ Liu, D.\ J.\ Scalapino
  and S.-C.\ Zhang, Phys. Rev. B{\bf 77}, 220503(R) (2008).

\bibitem{chen} W.-Q.\ Chen, K.-Y. Yang, Y. Zhou, and F.-C. Zhang,
 Phys. Rev. Lett. {\bf 102}, 047006 (2009).

\bibitem{kors} M.\ M.\ Korshunov and I.\ Eremin, Europhys. Lett. {\bf
  83}, 67003 (2008).

\bibitem{chub} A.\ V.\ Chubukov, arXiv:0902.4188v1.

\bibitem{varm} C.\ M.\ Varma, P.\ B.\ Littlewood, S.\ Schmitt-Rink, E.\
Abrahams, and A.\ E.\ Ruckenstein, Phys. Rev. Lett. {\bf 63}, 1996 (1989).
  
\bibitem{maha} G.\ D.\ Mahan, {\it Many-Particle Physics}, Kluwer
  Academic (2000).

\bibitem{pals} G.\ Palsson and and G.\ Kotliar, Phys. Rev. Lett. {\bf
  80}, 4775 (1998).

\bibitem{voru} P.\ Voruganti, A.\ Golubentsev, and S.\ John,
  Phys. Rev. B{\bf 45}, 13945 (1992).

\bibitem{kohno} H.\ Kohno and Y.\ Yamada, Prog. Theor. Phys.
  {\bf 80}, 623 (1988).

\bibitem{haul} K.\ Haule, A.\ Rosch, J.\ Kroha, and P.\ W\"olfle,
Phys. Rev. B{\bf 68}, 155119 (2003).

\bibitem{meri} J.\ Merino and R.\ H.\ McKenzie, Phys. Rev. B{\bf 61}, 7996
(2000). 

\bibitem{kont} See the review by K.\ Kontani, Rep. Prog. Phys. {\bf 71}, 026501 (2008).

\end{thebibliography}
\end{document}